\documentclass[12pt,preprint]{aastex}
\newcommand{\bdv}[1]{\mbox{\boldmath$#1$}}

\def\au{{\rm au}}

\def\masyr{{\rm mas}\,{\rm yr}^{-1}}
\def\kpc{{\rm kpc}}
\def\mas{{\rm mas}}

\def\max{{\rm max}}
\def\min{{\rm min}}
\def\rel{{\rm rel}}

\def\e{{\rm E}}
\def\bpi{{\bdv\pi}}
\def\bmu{{\bdv\mu}}

\def\btheta{{\bdv\theta}}

\def\bu{{\bf u}}

\begin{document}
\title{Path to Black Hole Astrometric Microlensing from 10-Year KMTNet Database }
%$M=17\,M_\odot$ and $v_\perp=120\,\kms$}

\author{\textsc{
Andrew Gould$^{1}$
}}

%----------------------------------------------------------------
\affil{$^{1}$Department of Astronomy, Ohio State University, 140 W. 18th Ave., Columbus, OH 43210, USA}

%\affil{$^{2}$Max-Planck-Institute for Astronomy, K\"{o}nigstuhl 17, 69117 Heidelberg, Germany}

%Andrew Gould (Ohio State University)

\begin{abstract}
  I present a Fisher analysis of the astrometric signature of black
  holes (BHs) that may lie in the 10-year KMTNet microlensing
  database, motivated by the recent success of the pilot study by
  \citet{segev26}, which reduced the scatter at $I=18$ to $\sigma\sim
  10\,\mas$. Seeing-limited ground-based observations still
    suffer from a great deal of systematics, but under the assumption
    that these systematics can be controlled, I find that with this
  nominal precision, the Einstein radius can be measured with errors
  $\sigma(\theta_\e)\sim 0.5\,\mas$ in the most intensively monitored
  fields ($\sim 12\, {\rm deg}^2$), and hence to twice that value in
  an additional $\sim 28\, {\rm deg}^2$ area.  This precision hardly
  depends on whether the peak of the event is during the observing
  season or not, as long as the impact parameter is within the broad
  range that is accessible to photometric recognition of the event,
  i.e., $u_0\la 1.5$.  However, longer events do have somewhat better
  precision.  The analysis presented here ignores blending, but this is
    a good first approximation for most sources $I\la 18$, which lie near
    or on the giant branch.
  I outline a practical program, combining photometric and
  astrometric searches, aimed at identifying isolated BHs in the
  KMTNet database and measuring their masses and distances.
  \end{abstract}

\keywords{gravitational lensing: micro}

\section{{Introduction}
\label{sec:intro}}

Microlensing is the only known method for detecting isolated black
holes (BHs).  Yet, a quarter century after \citet{gould00a} predicted
that of order 1\% of microlensing events (which by now number $\ga
50,000$) are caused by BHs, only one has been unambiguously
identified, namely OGLE-2011-BLG-0462
\citep{ob110462a,ob110462b,ob110462c}.  The origin of this mismatch
between potential and reality is that the only way to determine the
mass of a completely dark isolated object (and hence to confirm its
identity), is to simultaneously measure two quantities, the angular
Einstein radius, $\theta_\e$,
\begin{equation}
  \theta_\e\equiv\sqrt{\kappa M \pi_\rel};
  \qquad \kappa \equiv {4 G\over c^2\au}\simeq 8.144\,{\mas\over M_\odot}
  \label{eqn:thetaedef}
\end{equation}
and the microlensing parallax vector $\bpi_\e$ \citep{gould92,gould00b,gould04},
\begin{equation}
  \bpi_\e\equiv \pi_\e{\bmu_\rel\over\mu_\rel}; \qquad
      \pi_\e \equiv{\pi_\rel\over\theta_\e} = \sqrt{\pi_\rel\over\kappa M},
  \label{eqn:piedef}
\end{equation}
where $M$ is the mass of the lens, and $(\pi_\rel,\bmu_\rel)$ are the
lens-source relative parallax and proper motion.  Then, the mass and
distance are given by
\begin{equation}
  M = {\theta_\e\over \kappa \pi_\e}\qquad
  D_l = {\au\over \theta_\e\pi_\e + \pi_s},
  \label{eqn:massdist}
\end{equation}
where $\pi_s$ is the parallax of the lensed source.

Both measurements can be challenging.  However, the microlens parallax
can be measured directly from the same photometric light curve by which
microlensing events are discovered, particularly for long events
(see \citealt{gouldhorne} for a didactic explanation).  The main
difficulty is that, while BHs do indeed mainly have long Einstein timescales
(i.e., Einstein-radius crossing times),
\begin{equation}
  t_\e\equiv {\theta_\e\over \mu_\rel} = {\sqrt{\kappa M\pi_\rel}\over \mu_\rel},
  \label{eqn:tedef}
\end{equation}
due to their large mass, this same large mass tends to give them small
(hence, difficult to accurately measure) $\pi_\e$, according to
Equation~(\ref{eqn:piedef}).  This is particularly true for BHs in the
Galactic bulge, which (almost by definition) have small $\pi_\rel$.

These challenges are illustrated by the one successful BH mass measurement:
despite the fact that the lens was unusually nearby ($D_L\sim 1.5\,\kpc$)
and the BH was not exceptionally massive ($M\sim 8\,M_\odot$), the consequent
relatively ``large'' parallax $\pi_\e\sim 0.095$ required considerable
effort for an accurate measurement \citep{ob110462c}.

In the current context, the main takeaway point regarding $\pi_\e$ is that
we can often (though not always) expect to obtain good measurements with
sufficient effort, but we cannot expect some machine to just churn out
reliable annual parallax measurements from pipeline reductions of
  hundreds of events from ground-based survey data\footnote{Though routine
parallax measurements have already been achieved by combining satellite
data with ground-based survey data \citep{rybicki24}.}.

However, the key stumbling block to BH mass measurements has been, and
will continue to be, measurement of $\theta_\e$.  Of the 100+ microlensing
events with $\theta_\e$ measurements, the overwhelming majority
have been via finite-source effects, wherein the light curve is distorted
when the source star passes directly over a caustic in the lens structure
\citep{gould94a}.  The great majority of these caustic structures have
been due to binary lenses, in particular for planetary events because
the caustics are large, of the same order as the planetary perturbations
that make the event publishable as a ``planet discovery.''  By contrast,
for isolated lenses (the main subject of interest here), the caustic
consists of a single point, so the probability of such a crossing is
$p\simeq \rho \equiv \theta_*/\theta_\e$, i.e., the ratio of the
source radius to the Einstein radius.  For BHs, with their typically
large $\theta_\e$, $\rho\la 10^{-3}$.

\begin{figure}[htbp]
    \centering
  \includegraphics[width=0.8\linewidth]{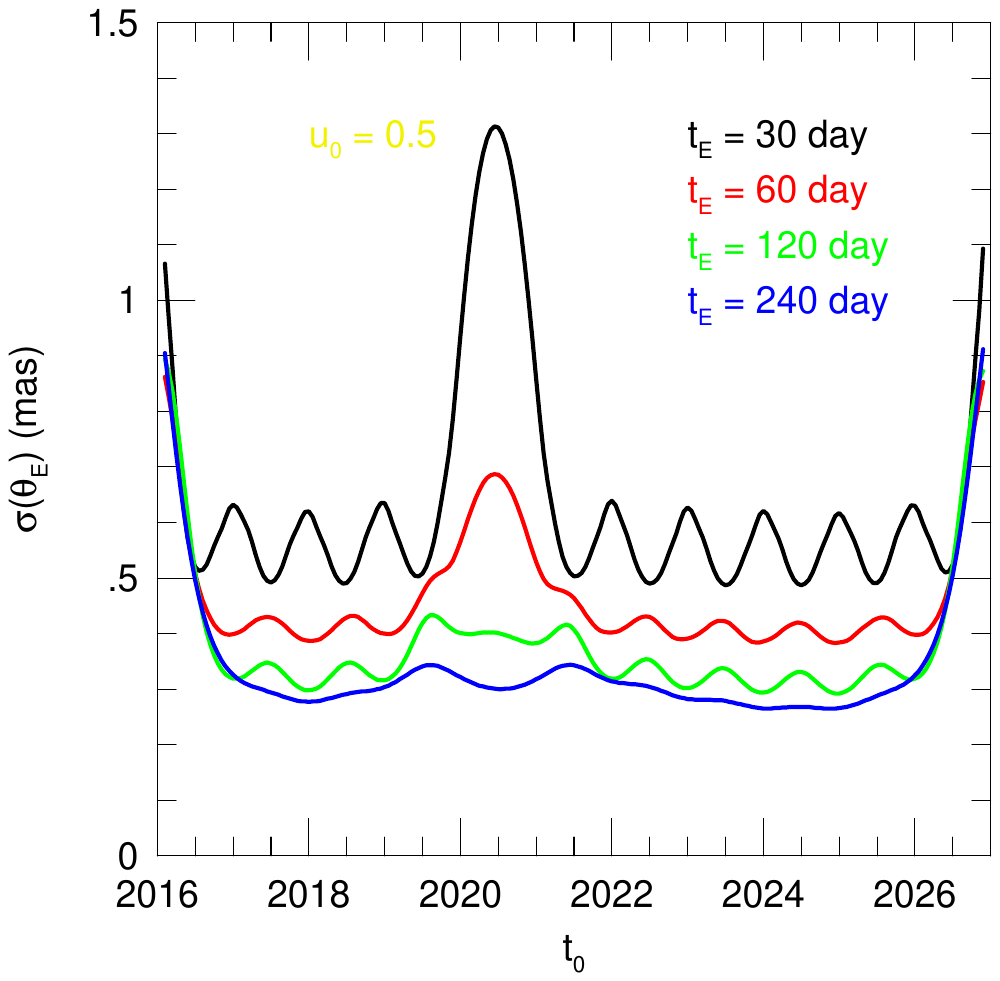}
  \caption{Einstein radius error estimates, $\sigma(\theta_\e)$ for
    events that peak any time during the KMTNet database of
    observation, 2016-2026, under the assumption of individual measurement
      precision of $\sigma=10\,$mas.  In particular, the events can peak any
    time of year.  A broad range of Einstein timescales, $t_\e$, is
    considered, but (so far) all simulated events have the same impact
    parameter, $u_0=0.5$.  Perhaps surprisingly, all simulated events
    have similar errors, except for those near the edge of the
    observation window and those during the Covid-19 year of 2020,
    during which observations were suspended.  }
\label{fig:longterm}
\end{figure}

Hence, other methods have been developed.  The first is
astrometric microlensing \citep{walker95,hnp95,my95} using high resolution
facilities such as the {\it Hubble Space Telescope (HST)},
which indeed was the approach of \citet{ob110462a} and \citet{ob110462b}.
In this method,
the centroid of the combined light of the two microlensed images is
displaced from the source by
\begin{equation}
  \Delta\btheta_{\rm cent}\equiv {\btheta_{\rm cent} -\btheta_s}
  = {\bu\over u^2+2}\theta_\e;
  \qquad \bu\equiv {\btheta_s -\btheta_l\over\theta_\e},
  \label{eqn:astulens}
\end{equation}
where $\btheta_s$ and $\btheta_l$ are the angular positions of the source
and lens, respectively.  Note that this achieves its maximum of
$\Delta\theta_{\rm cent} = \theta_\e/\sqrt{8}$ at $u=\pm\sqrt{2}$.
As I will explain in Section~\ref{sec:math}, this parameter must
be disentangled from four others, which generally requires a
long time series of such measurements.  In the case of OGLE-2011-BLG-0462,
these lasted a decade.  A secondary disadvantage of this method is that
candidate BH events are usually only recognized near (or after) peak, while
the astrometric measurement would usually be greatly improved by starting
the time series well before peak.  See Figure~\ref{fig:longterm}, below.

A second approach, which is potentially far more efficient, is
interferometric resolution of the two images \citep{delplancke01,dong19}.
In principle, this can yield a measurement of $\theta_\e$ from a single
observation:
\begin{equation}
  \theta_\e = {\Delta\theta_{\rm images}\over \sqrt{u^2 +4}},
  \label{eqn:imagesep}
\end{equation}
where $\Delta\theta_{\rm images}$ is the image separation and
$u$ is determined from the photometric light curve.  In practice, a single
measurement can yield degeneracies in $\Delta\theta_{\rm images}$, but these
can be resolved by a second epoch of observation, e.g., taken a month later.
The reason for the (again) quarter
century delay between the original \citet{delplancke01} proposal and its
first implementation, is that the Very Large Telescope Interferometer (VLTI)
required a series of upgrades to make it sensitive to the 
faint sources that give rise to the overwhelming majority of microlensing
events.

Although this interferometric
method will not be a major focus of the present work, we
do pause to mention an important subordinate advance that made it much
more practical, which will play a major role here. Traditionally, BH candidates
were identified solely on the basis of their long timescales,
$t_\e\propto M^{1/2}$.  But, as one can see from Equation~(\ref{eqn:tedef}),
long $t_\e$ can also be produced by big $\pi_\rel$ (nearby lenses) and/or
slow $\mu_\rel$.  Indeed \citet{kb162052} showed that such effects are
responsible for the overwhelming majority of long events, not high lens mass.
Hence, unless this problem were solved, the overwhelming majority of
very competitive VLTI observations would be ``wasted'' on contaminants.

Therefore, \citet{gould23} pointed out that an early parallax measurement
could greatly constrain the nature of the lens via the purely
mathematical relation
\begin{equation}
  \mu_\rel = {\theta_\e\over t_\e} = {\kappa M\pi_\e\over t_\e} = 1.5\,\masyr
  \biggl({\pi_\e\over 0.025}\biggr)
  \biggl({t_\e\over 50\,{\rm day}}\biggr)^{-1}
  \biggl({M\over M_\odot}\biggr).
\label{eqn:murelmass}
\end{equation}
Thus, for an event with a measured $\pi_\e=0.025$, one could set a minimum
timescale of $t_{\e,\min}=50\,$days for VLTI observations.  This would
eliminate the overwhelming majority of stellar contaminants while rejecting
very few BHs.  Specifically, for a given $\mu\la\sigma/2$, where
$\sigma = 2.9\,\masyr$ is the dispersion of bulge stars, the fraction
of events with $\mu_\rel<\mu$ is
$p=(\mu/\sigma)^3/6\sqrt{\pi} = 1.3\%(\mu/1.5\,\masyr)^3$ \citep{masada}.
Hence, 98.7\% of all events generated by $M=1\,M_\odot$ stars would be
eliminated, as would 99.8\% of all stars with a more typical mass of
$M=0.5\,M_\odot$.  On the other hand, $10\,M_\odot$ BHs would be eliminated
only if they were moving extremely fast, $\mu_\rel>15\,\masyr$.
This led \citet{gould23} to a suggested selection criterion:
\begin{equation}
t_\e > 50\,{\rm days}\biggl({\pi_\e\over 0.025}\biggr).
\label{eqn:critgould}
\end{equation}
Now, \citet{gould23} realized that accurate $\pi_\e$ measurements
that were based on survey data and were near
enough to peak to trigger VLTI measurements, would be extremely rare.
Hence, he advocated a parallax satellite \citep{refsdal66,gould94b,yee15},
which would be capable of such rapid measurements.  Subsequently,
\citet{wu26} found a way to modify the \citet{gould23} approach
that would evade the parallax-satellite requirement, and so the
time, money and uncertainty that this would entail.  However, this
modification is not germane to the present work, which focuses on
archival events.

An alternative to these two approaches was recently presented by
\citet{segev26}:  Extract $\theta_\e$ from astrometric
microlensing measurements using the 10-year, $\sim 100\,$deg$^2$
ground-based, seeing-limited, KMTNet database \citep{kmtnet}.
Their approach was to obtain astrometric time series for a large
ensemble of stars, which would enable one to search for BH candidates
independent of the photometric light curves.  In their inaugural paper,
they specifically applied their
algorithm to KMTC (CTIO, Chile) data, and their Figure~6 shows that
they were able to obtain scatter of $\sigma\sim 10\,\mas$ for a
substantial fraction of stars at $I=18\,$mag in the relatively
uncrowded KMT field BLG17.  See Figure~12 of \citet{eventfinder} for the
KMT field map.  For typical KMT field extinctions, $A_I\sim 2$, this would
be a low luminosity giant.  While certainly less common that the turnoff
star sources that dominate microlensing statistics, such stars are not rare.
And this magnitude is similar to what is accessible using the two other
techniques.

Inspired by this work, I lay the foundation for a complementary approach:
first, search the KMT photometric database for BH candidates using the
\citet{gould23} criterion, Equation~(\ref{eqn:critgould}), based on $t_\e$
and $\pi_\e$; second,
evaluate what astrometric precision would be required to reliably determine
the mass, i.e., to measure,
\begin{equation}
  \theta_\e=\kappa M_{\rm bh}\pi_\e=4.1\,\mas{M_{\rm bh}\over 10\,M_\odot}
        {\pi_\e\over 0.05};
  \label{eqn:thetaebh}
\end{equation}
and finally, to apply (and possibly fine tune) the \citet{segev26} approach
to optimize the astrometry of this {\it particular} microlensed source
to hopefully reach the required precision.

\section{{Formalism}
\label{sec:math}}

The first step is to estimate the precision of a $\theta_\e$ measurement,
given the specific epochs of the time series of the data, their individual
precision, and the parameters that describe the photometric light curve,
which in principle consist of $(t_0,u_0,t_\e,\bpi_\e)$, that is the
time of peak, the impact parameter (in units of $\theta_\e$), the
Einstein timescale, and the microlens parallax vector.  In addition,
one should specify the event coordinates, which imply different cadences
(see Figure~12 of \citealt{eventfinder}) and somewhat different
windows of observation.  This would seem to be of sufficient complexity
and diversity as to require a Monte Carlo approach.  However, I will show
immediately below and in Section~\ref{sec:fisher}
that a much simpler, semi-analytic approach is possible.

The first point is that in this study it is appropriate to ignore $\bpi_\e$,
that is, effectively to set $\bpi_\e=0$.  The effect of including it would be
to induce annual oscillations in the normalized source-lens separation, of
amplitude $\Delta u=\pi_\rel/\theta_\e=\pi_\e$, which for most BHs is
a few percent.

Second, the observational pattern is basically the same each year and
is set by the bulge observing window.  This window shifts slightly to
earlier times as one goes from east to west.  But we will see that
much larger changes in the peak time of the event relative to the
observing window have only a minor effect.  Hence, I will fix the
observing pattern.  For the roughly 12 deg$^2$ for which the nominal
cadence is $\Gamma=4\,{\rm hr}^{-1}$, I find that the observing rate is constant
at 40/day for 40 days around the summer solstice, and then falls
linearly to 6/day at 105 days before this peak interval and also
to 6/day at 95 days after.  That is, there is a 105+40+95=240 day observing
season, with a nominal scheduled number of observations of 6200/year.
However, the actual number of observations (due to weather, equipment
and technical problems, etc.) is typically 4800.  Because KMT takes
data under all conditions, including e.g., all lunar phases, I further
scale back the assumed rate to 4000/year as my estimate of the annual
rate of usable observations.  I set the precision of each observation
at $\sigma = 10\,\mas$.

Then, to interpret the result for other fields, with other values of
$\Gamma$ (see Figure~12 of \citealt{eventfinder}), and for events
for which one expects larger or smaller $\sigma$, one should just
multiply the results derived below by 
\begin{equation}
240  \sigma(\theta_\e) \rightarrow
 \sigma_{\rm nominal}(\theta_\e){\sigma\over 10\,\mas}\sqrt{4/{\rm hr}\over\Gamma}.
  \label{eqn:rescale}
\end{equation}

\subsection{{Fisher Analysis}
\label{sec:fisher}}

We can write the apparent source position, $\btheta_{s,\rm app}(t)$ as the sum of
its true position, $\btheta_{s}(t)=\btheta_{s,0}+\bmu_s(t-t_0)$ and another
term that represents the displacement of the image centroid from the source
position:
\begin{equation}
\btheta_{s,\rm app}(t) = \btheta_{s,0}+\bmu_s\tau{t_\e\over \rm yr}
+ {\bu\over u^2 +2}\theta_\e; 
\qquad
\bu \equiv (\tau\cos\psi-u_0\sin\psi,\tau\sin\psi +u_0\cos\psi),
\label{eqn:appsource}
\end{equation}
where $\tau\equiv (t-t_0)/t_\e$. % and $\beta\equiv u_0$.
This a function of six parameters, effectively
$(\btheta_{s,0},\bmu_s,\bmu_\rel)$ because $\theta_\e/t_\e = |\bmu_\rel|$
and $\psi$ represents the direction of $\bmu_\rel$.  In principle,
we could suppress $\psi$ (set $\psi\equiv 0$) because this direction is
(or should be) the same as the direction of $\bpi_\e$, which is known
from the light-curve parallax measurement.  However, incorporating it as a
free parameter permits an independent check on the direction of $\bpi_\e$.
Moreover, within the context of the Fisher analysis, and assuming
  isotropic measurement errors as is done here,
one may prove that
the correlation between $\theta_\e$ and $\psi$ is identically zero,
so including it has no effect of the error estimate.  In practice,
  the errors are unlikely to be perfectly isotropic due, e.g., to the
  fact that the parallactic angle changes over the night and season, but
  we only expect the Fisher analysis to give overall guidance and not
  to reflect every practical detail.

Thus, there are six parameters $(\btheta_{s,0},\bmu_s,\theta_\e,\psi)$
and two measurements $(\theta_{s,x,\rm app}(t_i),\theta_{s,y,\rm app}(t_i))$
for each observations.  Hence, we estimate the error in $\theta_\e$
by linearizing Equation~(\ref{eqn:appsource}), i.e., evaluating
six linearized ``trial functions'', as the derivatives of the
six parameters, $a_i$,  with respect to the two measured quantities
$(f_{i,x},f_{i,y}) = \partial \btheta_{s,\rm app}/\partial a_i$.
Explicitly,
\begin{equation}
  f_{1,x} = f_{2,y} = 1;\quad
  f_{3,x} = f_{4,y} = \tau{t_\e\over \rm yr};\quad
  [(f_{5,x},f_{5,y}),(f_{6,x},f_{6,y})]
  = {[(\tau,u_0),(-u_0,\tau)\theta_\e]\over \tau^2+u_0^2 +2},
  \label{eqn:trial}
\end{equation}
with the remaining four vanishing,
$f_{2,x}=f_{1,y}=f_{4,x}=f_{3,y}\equiv 0$.
Then we evaluate the inverse covariance matrix\footnote{The form of
 Equation~(\ref{eqn:bmat}) implicitly assumes that the errors $\sigma$
  of the individual data points are uncorrelated.  See \citet{gould03}
  (top of page 2) for the more general form.  Figure~9 of \citet{segev26}
  shows a roughly flat power spectrum of the astrometric errors, which
  gives qualitative support from this pilot study that correlations among 
  the errors can be kept to a minimum in practice.},
\begin{equation}
  b_{ij} = \sum_k {f_{i,x}(t_k)f_{j,x}(t_k)\over \sigma^2}
 + \sum_k {f_{i,y}(t_k)f_{j,y}(t_k)\over \sigma^2},
  \label{eqn:bmat}
\end{equation}
invert to form the covariance matrix, $c= b^{-1}$, and finally evaluate the
error in $\theta_\e$,
\begin{equation}
  \sigma(\theta_\e) = \sqrt{c_{5,5}}.
  \label{eqn:sigmathetae}
\end{equation}

The analysis presented here ignores blending, which is a reasonable
  first approximation for sources at $I\la 18$, i.e., those for which
  the \citet{segev26} pilot study achieved good results (see their Figure~6).
  This is because, at typical extinctions $A_I\sim 2$, such stars lie
  $\Delta I\la 1.5$ below the clump, i.e., they are giants.  Because
  of the low surface density of stars of comparable brightness (or even
  several times fainter), experience shows that such sources are
  typically at most weakly blended.  Nevertheless, in any individual case,
  blending can be a moderate, or even severe, problem, and it must be
  carefully checked.

\section{{Application to the KMTNet Database}
\label{sec:app}}

\subsection{{Variations in $\sigma(\theta_\e)$ Over 10 Years}
\label{sec:long}}

With this machinery in place, we can now investigate the errors in $\theta_\e$
over several ranges of parameters.  We first consider events occurring at
any time during the 10 years of observations, 2016-2026 (with a one year hiatus
in 2020 due to Covid-19).  For this initial investigation, we restrict
consideration to a single representative value of $u_0=0.5$, but consider
a broad range of event timescales, $30\,{\rm day}\leq t_\e \leq 240\,{\rm day}$.
The results are shown in Figure~\ref{fig:longterm}.

Let us focus first on the red ($t_\e=60\,$day) curve, which is a typical
duration for a BH candidate.  The most notable feature is that, apart from
the 2020 Covid-19 year and the edges of the observations at 2016.0 and 2027.0,
the curve is nearly flat.  That is, regardless of what time of year the
event peaked, $\sigma(\theta_\e)\sim 0.4\,\mas$.  This may appear
counter-intuitive.  We will examine the origins of this effect in
Section~\ref{sec:ellipse}.
Even for events peaking during 2020, when we are assuming zero observations,
the errors are not dramatically worse.  And, if we ignore 2020, then the
variations in $\sigma(\theta_\e)$ are quite moderate even for $t_\e=30\,$day
(black curve).

\subsection{{Variations in $\sigma(\theta_\e)$ with Respect to $u_0$}
\label{sec:beta}}

\begin{figure}[htbp]
    \centering
  \includegraphics[width=0.8\linewidth]{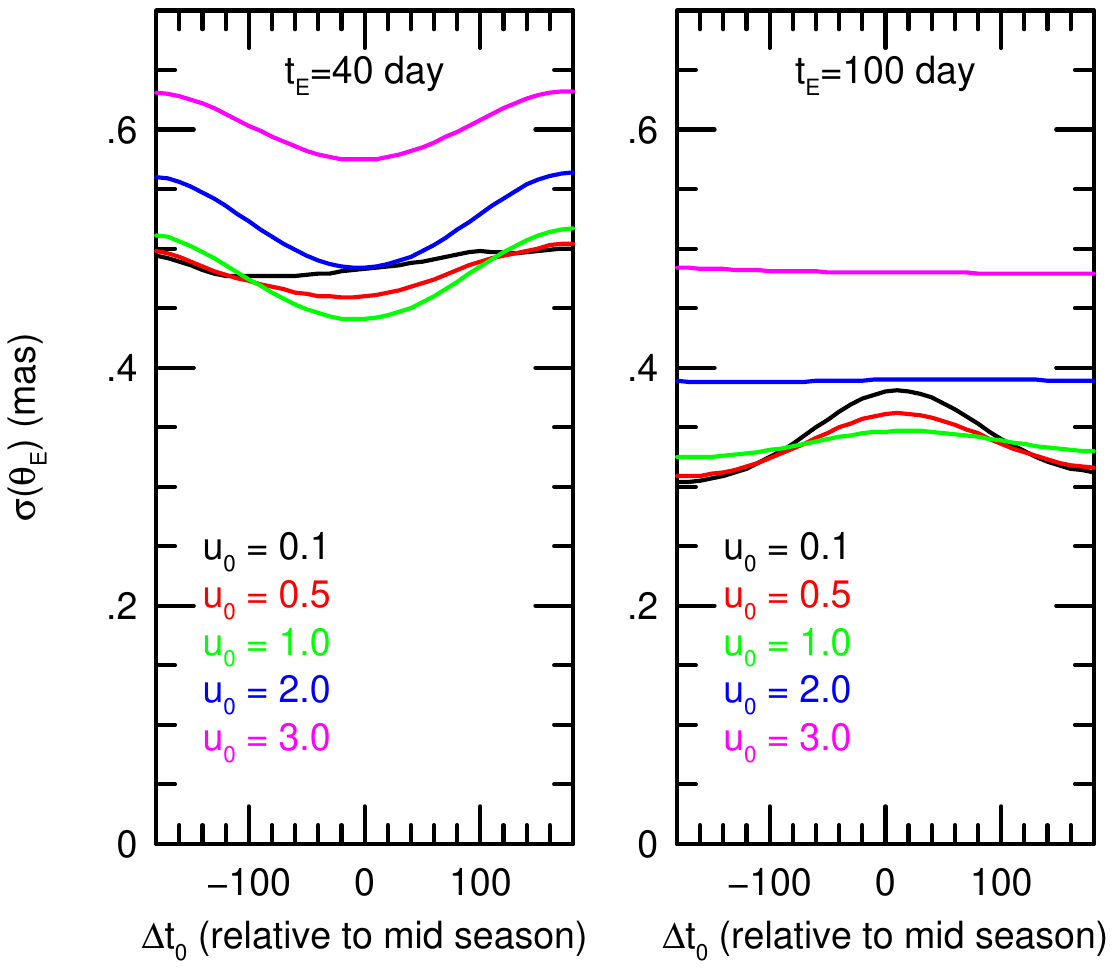}
  \caption{Einstein radius error estimates, $\sigma(\theta_\e)$ for
    events that peak any time during 2023 and with a variety of
    impact parameters $0.1\leq u_0\leq 3.0$.  Two Einstein timescales
    $t_\e$ are considered: 40 day (left) and 100 day (right).  For
    events that can plausibly be recognized photometrically, i.e.,
    $u_0\la 1.5$, the errors do not vary much with $u_0$.  However,
    there are modest ``dips'' near $\Delta t_0=0$ in the left panel
    and are modest ``bumps'' near $\Delta t_0=0$ in the right panel.
    These are explained in Figures~\ref{fig:ell40} and \ref{fig:ell100},
    respectively.
    }
\label{fig:bothdbeta}
\end{figure}

Figure~\ref{fig:bothdbeta} shows $\sigma(\theta_\e)$ for two timescales,
$t_\e=40\,$day and $t_\e=100\,$day, in the two panels.  The events peak
over the full range of dates within 2023.  A large range of impact parameters
are explored, $0.1\leq u_0\leq 3.0$.  The main takeaway from this
figure is that for fixed $t_\e$ and for events that can plausibly be
recognized photometrically, $u_0\la 1.5$, the precision of the
$\theta_\e$ measurement does not qualitatively depend on either $u_0$
or the time of year when the event peaks.  That is, the lesson from
Figure~\ref{fig:longterm}, i.e., that the main systematic dependence
of $\sigma(\theta_\e)$ is a weak trend toward improvement with
increasing $t_\e$, applies to all $u_0\la 1.5$, not just $u_0=0.5$.
This should not be surprising, given that the maximum astrometric deviation
  (from Equation~(\ref{eqn:astulens})) is $\theta_\e/\sqrt{8}$
  at $u=\sqrt{2}$, so that all
events with $u_0\leq \sqrt{2}$ experience this maximum deviation.

Secondarily, Figure~\ref{fig:bothdbeta} shows weak trends $\sigma(\theta_\e)$
with $\Delta t_0$.  In particular, there are ``dips'' near
$\Delta t_0\sim 0$ for $t_\e = 40\,$day and ``bumps'' near
$\Delta t_0\sim 0$ for $t_\e = 100\,$day.  These ``dips'' and ``bumps''
are substantially smaller than the factor $\sim 2$ difference in
individual data-point precision between $I=17$ and $I=18$ sources
in Figure~6 of \citet{segev26}, so they are not a dominant consideration
in an overall sense.  Nevertheless, it is curious why they occur, and
why they have opposite signs in the two cases.  I will address these questions
in Section~\ref{sec:ellipse}.

Finally, the $u_0=3.0$ (magenta) curves are included mainly for
illustration.  Such events, with maximum magnification $A_\max=1.017$,
would rarely, if ever, be recognized photometrically in current
ground-based microlensing experiments.  Nevertheless, these curves
illustrate that astrometric sensitivity declines only slowly as the
events leave the photometric domain, which raises the possibility
(which will not be addressed here) of a purely astrometric search
for large $\theta_\e$ events.

\subsection{Populating the Astrometric Ellipse}
\label{sec:ellipse}

\begin{figure}[htbp]
    \centering
  \includegraphics[width=0.8\linewidth]{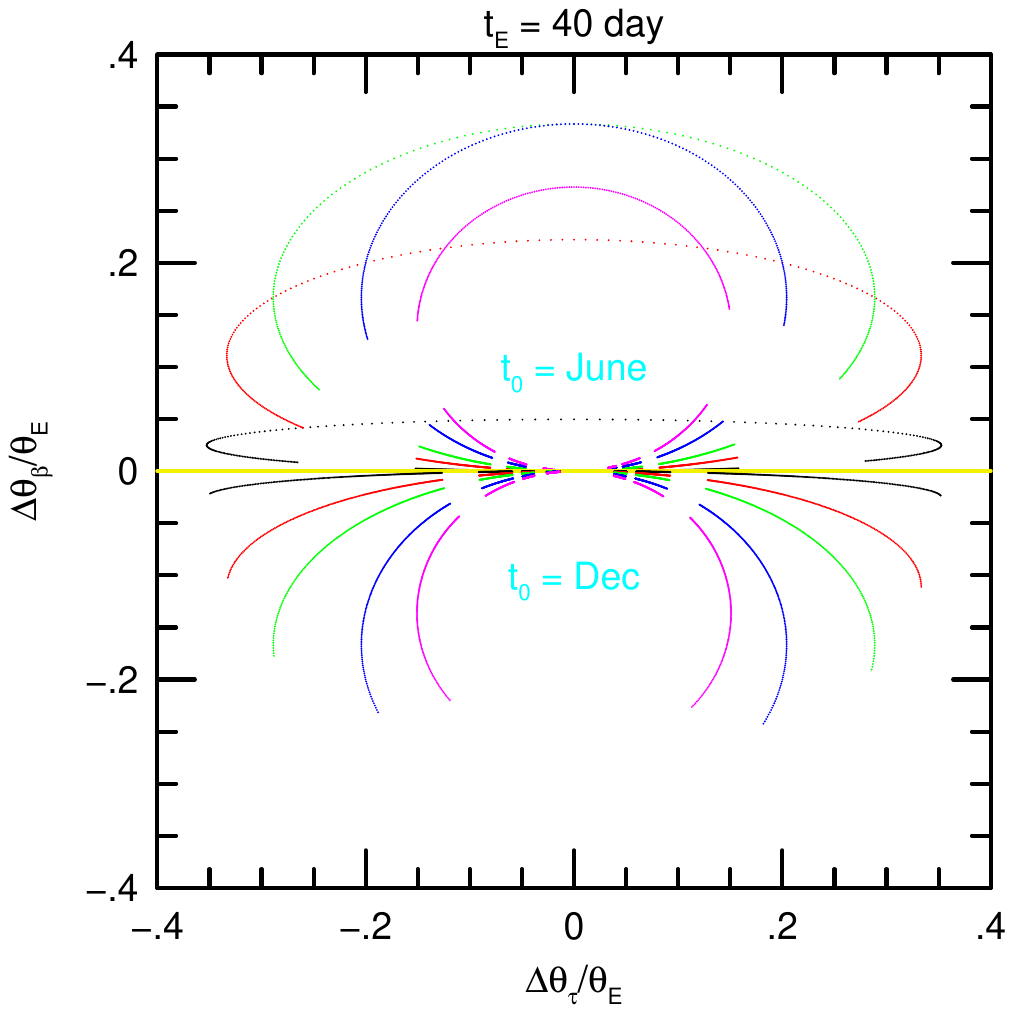}
  \caption{``Astrometric ellipses'' for $t_\e=40\,$day events for
    five values of $u_0$ (color-coded as in Figure~\ref{fig:bothdbeta}),
    and for two values of $t_0$: summer solstice (upper half of figure)
    and winter solstice (lower half of figure). The ``dips'' in the
    left panel of Figure~\ref{fig:bothdbeta} in the red and green curves
    are caused by better coverage of the left-right wings of the
    ellipse for the summer solstice ellipses.  Regarding the ``dips''
    in the blue and magenta curves, these are caused by the better
    coverage of the summer solstice ellipses near the extreme vertical
    regions.
    }
\label{fig:ell40}
\end{figure}

Figure~\ref{fig:ell40} shows how the $t_\e=40\,$ day
``astrometric ellipses'' are populated by observations for two values of
$t_0$ i.e., summer solstice (upper half) and winter solstice (lower half),
and five different values of $u_0$, which are color-coded as in
Figure~\ref{fig:bothdbeta}.  The reason for the ``dips'' for all $u_0>0.1$
in the $t_\e=40\,$day panel of that Figure can easily be seen.
For example, for the summer solstice, the red and green
ellipses are all well populated  at the right and left edges, whereas
the winter solstice ellipses of the same color are not.  For the cases
of the green and magenta ellipses, which are more circular, they are
more populated at the extreme upper end for the summer solstice than they are in
the corresponding extreme lower end for the winter solstice.

\begin{figure}[htbp]
    \centering
  \includegraphics[width=0.8\linewidth]{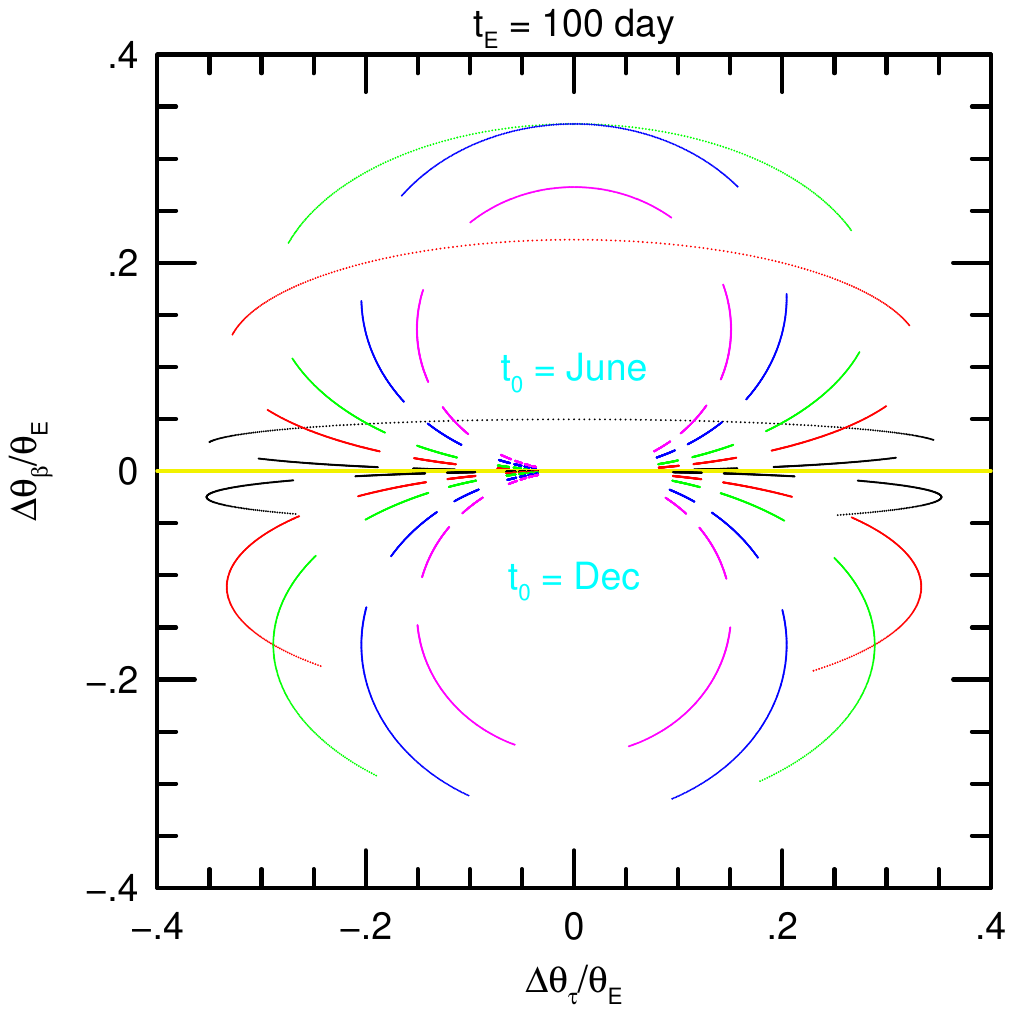}
  \caption{Same as Figure~\ref{fig:ell40} except for $t_\e=100\,$day.
    The ``bumps'' in the
    right panel of Figure~\ref{fig:bothdbeta} in the black and red curves
    are caused by better coverage of the left-right wings of the
    ellipse for the winter solstice ellipses.  
    }
\label{fig:ell100}
\end{figure}

Figure~\ref{fig:ell100} shows the same ``astrometric ellipses'' for
$t_\e=100\,$day.  In this case, the right and left edges of the
black and red ellipses are better populated for the winter solstice than for
the summer solstice, and this explains why there are ``bumps'' in these two
colors in the $t_\e=100\,$day panel of Figure~\ref{fig:bothdbeta}.

\section{{Implications for Black Hole Searches}
\label{sec:imply}}

What is most striking about the results reported in
Section~\ref{sec:app} is how boring they are.  Broadly, there is
little variation in the precision of the $\theta_\e$ measurement by
$t_0$ (time of peak) or $u_0$ (impact parameter, so maximum
magnification).  Precision generally improves for longer timescale
($t_\e$) events, but not dramatically.  Of course, the precision will
vary considerably due source brightness (so underlying astrometric
measurement precision), whose effect can be roughly incorporated by a
simple multiplicative factor.  And it will also be affected by field cadence,
$\Gamma$, but this can also be incorporated by a multiplicative factor
$\sqrt{(4/{\rm hr})/\Gamma}$.
And it will also vary based on local
conditions, e.g., crowding, blending, ambient diffraction spikes,
etc., which must be evaluated on a case-by-case basis.  Finally, there
  is undoubtedly some floor on the precision that can be achieved on
  individual events, perhaps rooted in stellar density.  But this
  can only be identified and quantified based on practical work to
  reduce systematics to a minimum in individual cases.

What this tells us is about a BH search through archival KMT data is that
it must begin with a search through {\it photometric} light curves
for suitable candidates, which can then be investigated individually
for an astrometric signal.  That is, prior to such a photometric investigation,
there is no characteristic of one event (other than source brightness)
that makes it a better astrometric candidate than another.

The key roadblock to a photometric search for candidates is that it requires
(or at least would strongly benefit from) tender loving care (TLC)
re-reduction of the data, as developed by \cite{yang24}.  While this
is straightforward for any particular event, it has only been applied
on an ``industrial scale'' to one full year of KMT events, namely 2023.

Hence, one could start with events from that year to identify those that
approximately satisfy the \citet{gould23} criterion, i.e.,
Equation~(\ref{eqn:critgould}).  For each such case of approximate
satisfaction, one could obtain TLC for the 2022 and 2024 data, and
combine these with the 2023 data for the same event, in order to make
a more precise assessment.

For other years, one could focus TLC reduction efforts on events with
a ``reasonable chance'' of being good candidates, for example those
with pipeline reduction parameters $t_\e>t_{\e,\rm lim}$ and $I_s<I_{s,\rm lim}$.
For example, for 2025, and $t_{\e,\rm lim}=30\,$day and $I_{s,\rm lim}=19$,
I find 241 events that would require TLC out of a total of 3348.
For $t_{\e,\rm lim}=40\,$day and $I_{s,\rm lim}=19$,
I find 174.

%\begin{equation}
%  \label{eqn:}
%\end{equation}

\acknowledgments

I thank Eran Ofek and Yossi Shvartzvald for valuable discussions.

%
%This research has made use of the KMTNet system operated by the Korea Astronomy and Space Science Institute (KASI) and the data were obtained at three host sites of CTIO in Chile, SAAO in South Africa, and SSO in Australia.
%
%Work by C.H. was supported by the grant (2017R1A4A101517) of National Research Foundation of Korea.
%
%The OGLE project has received funding from the National Science Centre, Poland, grant MAESTRO 2014/14/A/ST9/00121 to AU.
%
%The MOA project is supported by JSPS KAKENHI Grant Number JSPS24253004, JSPS262%47023, JSPS23340064, JSPS15H00781, and JP16H06287.
%

%

%\input tab

%\input tabparm
%\input tab2

%\input tab3

%\input tabnames

%\input tab1253

%\input tabcmd

%\begin{figure}
%\plotone{../kb1253/KB211253_L1.eps}
%\caption{Light curve and model for KMT-2021-BLG-1253}
%\label{fig:1253lc}
%\end{figure}

\end{document}